\documentclass{article}
\usepackage{amsmath}
\usepackage{graphicx}
\usepackage{amssymb}
\usepackage{amsfonts}
\usepackage{geometry}
\usepackage{hyperref}
\usepackage{physics}
\usepackage{bm}
\usepackage{graphicx,color}
\begin{document}

\title{\bf Bianchi IX dynamics with a phantom field}

\author{Adel Awad${}^{1,2}$, Dmitry Chirkov${}^{3}$, Alexey Golovnev${}^{2}$, Alexey Toporensky${}^{3}$\\ \\
{\small ${}^{1}${\it 
Department of Physics, Faculty of Science, Ain Shams University, }}\\ 
{\small\it Cairo 11566, Egypt}\\
{\small ${}^{2}${\it Centre for Theoretical Physics, The British University in Egypt,}}\\ 
{\small\it P.O. Box 43, El Sherouk City, Cairo 11837, Egypt}\\
{\small ${}^{3}${\it Sternberg Astronomical Institute, Moscow State University, Moscow 119991, Russia}} \\ \\
{\small awad.adel@gmail.com}\\
{\small chirkovdmt@gmail.com}\\
{\small agolovnev@yandex.ru}\\
{\small atoporensky@gmail.com}
}
\date{}

\maketitle

\begin{abstract}
We consider Bianchi IX dynamics of a Universe filled with a massless phantom field. Such an exotic matter source enables regimes impossible in vacuum or with a standard scalar field. In particular, two Kasner indices of BKL oscillations can be simultaneously negative, and the absolute value of a negative index can be large. We describe the consequences of these features and explain the nature of volume oscillations recently discovered  in such a system by numerical methods.
\end{abstract}

\section{Introduction}
The Belinskii-Khalatnikov-Lifshitz (BKL) oscillations \cite{BKL} have been attracting a lot of attention ever since their discovery,
mainly because they are assumed to be the general regime of approaching the cosmological singularity. Some interesting mathematical results have been established, especially considering 
the chaotic properties of this regime. More precisely, the regime in question can be represented
by a sequence of Bianchi I Kasner epochs, and this discrete sequence (referred to as the Mixmaster map
\cite{Misner})
is shown to be chaotic \cite{Barrow1, Barrow2}, though a rigorous mathematical proof that the Mixmaster map is indeed an attractor for a general vacuum Bianchi IX dynamics is still absent (for a review of what is currently
known and what is still unproved, see for example \cite{Uggla}).

Apart from the vacuum case, the possible influence of an ordinary matter has been studied, starting from the work of the founders of BKL regime \cite{fields}. It is easy to show that in the vicinity of a cosmological singularity "matter does not matter" except for a stiff fluid\footnote{or a super stiff fluid if one is all right about rather exotic sources}, the most natural physical representation of which is a massless scalar field. The disappearance of chaos in the presence of a standard massless scalar field has been shown in \cite{fields}. Recent developments in cosmology, however, sometimes invoke exotic matter, so that the question about the fate of BKL oscillations in presence of some kind of non-standard matter arises naturally. Since it is a scalar field that modifies the vacuum solutions seriously, studies of a Bianchi IX dynamics with an exotic scalar field is a reasonable generalization of the standard BKL scenario. 

This program has been started rather recently with investigation of Bianchi IX Universe in Horndeski theory -- the most general theory of a scalar field which keeps the dynamical equations to be differential equations of the second order \cite{Horndeski}. In the paper \cite{Sushkov} it was claimed that numerical integration had shown non-monotonic behavior of the mean scale factor, i.e. a bounce. This means that the volume initially decreasing, while going to the past, turns to an increasing stage. This is in a sharp contrast to the usual BKL picture where the volume decreases linearly towards the singularity.

Arguably, the simplest case of an exotic matter is
a massless scalar field with a wrong sign of its kinetic term, a so-called phantom field. A special numerical study of Bianchi IX dynamics with a phantom field has been done in \cite{Volkov}, and volume oscillations have been observed. One of the goals of the present paper is to describe this phenomenon from a theoretical point of view.

One of interesting features of the BKL analysis with a phantom scalar field is that the parameter $q$ associated with the phantom scalar field is no longer bounded. This feature changes the Kasner exponents of the solution in two ways: first, it is possible to have two negative exponents; second, some exponents can take arbitrarily large values if $q$ is large enough.

The paper is organized as follows. In Sec.2 we remind the reader of the basic equations and properties of the standard BKL scenario. In Sec.3 we study the effect of two negative indices. In Sec.4 we consider indices unbounded from below. This allows us to explain the volume oscillations. Sec.5 gives a summary of the results obtained.

\section{Bianchi IX Cosmology with a Scalar Field: Field Equations}

We consider the anisotropic Bianchi IX cosmology, where the spatial section is a three-sphere $S^3\cong\mathrm{SU}(2)$, and we work in the basis of
left-invariant one-forms $\omega^a$ ($a=1,2,3$) satisfying
$$d\omega^a = \frac{1}{2}\epsilon^{abc}\,\omega^b\wedge\omega^c.$$
The one-forms $\omega^a$, in the classical notation of BKL, can be written in Euler angles $(\psi,\theta,\varphi)$ as
\begin{align}
  \ell &= (\sin\psi,\;-\cos\psi\sin\theta,\;0),\\
  m    &= (\cos\psi,\;\phantom{-}\sin\psi\sin\theta,\;0),\\
  n    &= (0,\;\cos\theta,\;1).
\end{align}
The most general diagonal Bianchi IX metric then reads

\begin{equation}
  ds^2 = -dt^2 + \gamma_{ij}\,dx^i dx^j
       = -dt^2
         + a^2(t)\,(\omega^1)^2
         + b^2(t)\,(\omega^2)^2
         + c^2(t)\,(\omega^3)^2,
  \label{eq:metric}
\end{equation}
where $$\gamma_{ij} = a^2\ell_i\ell_j + b^2 m_i m_j + c^2 n_i n_j,$$
and $a(t)$, $b(t)$, $c(t)$ are the three directional scale factors. We include a minimally coupled massless scalar field $\phi = \phi(t)$, with either positive or negative kinetic term!
 
\subsection{Field Equations}
 
The Einstein field equations together with the scalar field equation
yield the following system (dots denote $d/dt$):

\begin{subequations}
\label{eq:fieldeqs}
\begin{align}
  (\dot{a}bc)^{\centerdot}
  &= \frac{1}{2abc}\Bigl[(b^2-c^2)^2 - a^4\Bigr],
  \label{eq:fe_a}\\[4pt]
  (a\dot{b}c)^{\centerdot}
  &= \frac{1}{2abc}\Bigl[(a^2-c^2)^2 - b^4\Bigr],
  \label{eq:fe_b}\\[4pt]
  (ab\dot{c})^{\centerdot}
  &= \frac{1}{2abc}\Bigl[(a^2-b^2)^2 - c^4\Bigr],
  \label{eq:fe_c}
\end{align}
\end{subequations}
and the scalar field equation $\square\phi=0$,
\begin{equation}
  \dot\phi\!\left(\frac{\dot a}{a}+\frac{\dot b}{b}+\frac{\dot c}{c}\right)
  + \ddot\phi = 0,
  \label{eq:scalar_eom}
\end{equation}
together with its constraint equation
\begin{equation}
  \frac{\ddot a}{a} + \frac{\ddot b}{b} + \frac{\ddot c}{c}
  + \frac{\epsilon}{2}\dot\phi^2 = 0,
  \label{eq:constraint_t}
\end{equation}
where $\epsilon= +1$ for the usual scalar field, and $\epsilon= -1$ for the phantom field. One can find the accelerations, such as $\frac{\ddot a}{a}$, from the gravity equations (\ref{eq:fieldeqs}) and substitute them into the "constraint" equation (\ref{eq:constraint_t}) in order to get the Hamiltonian constraint, indeed.

In the physical time $t$, we can divide both sides of equations (\ref{eq:fieldeqs}) by the volume factor $abc$ and see that, as long as no dimension is stationary, and if all the scale factors are equally large, then the right-hand sides can be neglected and the evolution be approximated by a Bianchi I regime. As is well-known, a bounce to another Bianchi I regime would generically occur on the way to the cosmological singularity when one of the scale factors becomes large compared to others, so that its size can play a role. For describing the transition between two epochs, one commonly uses another time variable.
 
\subsection{Logarithmic Scale Factors and Conformal Time}
We introduce logarithmic scale factors $\alpha$, $\beta$, $\gamma$ as
\begin{equation}
  a = e^{\alpha(\tau)},\qquad
  b = e^{\beta(\tau)},\qquad
  c = e^{\gamma(\tau)},
\end{equation}
and pass to another time variable $\tau$ defined by
\begin{equation} 
\label{tau}
  \Lambda\,dt = abc\,d\tau
\end{equation}
where $\Lambda$ is an arbitrary constant. It would be enough to consider the case of $\Lambda=1$ which is somewhat analogous to the notion of conformal time. However, for clarity of discussion, it is convenient to have it in the general case.

Under this change of variables (\ref{tau}), the equations (\ref{eq:fieldeqs}) take the form
\begin{subequations}
\label{confteq}
\begin{align}
  2\Lambda^2\alpha_{,\tau\tau}
  &= (b^2-c^2)^2 - a^4
   = e^{4\beta}+e^{4\gamma}-2e^{2(\beta+\gamma)}-e^{4\alpha},\\
  2\Lambda^2\beta_{,\tau\tau}
  &= (c^2-a^2)^2 - b^4
   = e^{4\gamma}+e^{4\alpha}-2e^{2(\gamma+\alpha)}-e^{4\beta},\\
  2\Lambda^2\gamma_{,\tau\tau}
  &= (a^2-b^2)^2 - c^4
   = e^{4\alpha}+e^{4\beta}-2e^{2(\alpha+\beta)}-e^{4\gamma}.
\end{align}
\end{subequations}
The commas denote the derivatives:  $\alpha_{,\tau}\equiv d\alpha/d\tau$, etc.
The scalar field equation (\ref{eq:scalar_eom}) becomes
\begin{equation}
  \phi_{,\tau\tau} = 0, \label{phi}
\end{equation}
and its constraint (\ref{eq:constraint_t}) reads
\begin{equation}
  \alpha_{,\tau\tau}
  + \beta_{,\tau\tau}
  + \gamma_{,\tau\tau}
  + \frac{\epsilon}{2}\phi_{,\tau}^2 = 2 [\alpha_{,\tau} \beta_{,\tau}+\alpha_{,\tau} \gamma_{,\tau}+\gamma_{,\tau} \beta_{,\tau}]. \label{constraint}
\end{equation}

Equation (\ref{phi}) integrates immediately to
\begin{equation}
  \phi_{,\tau} = {\tilde q} = \mathrm{const},
  \label{eq:phi_const}
\end{equation}
where ${\tilde q}$ is the conserved scalar momentum.  This conservation law
holds exactly throughout the entire cosmological evolution. It is a direct
consequence of the shift symmetry, $\phi\to\phi+\mathrm{const}$. Note however that the value of ${\tilde q}$ depends on the chosen time variable $\tau$, while after a bounce we will rescale the time for better correspondence to the Bianchi I approximation parameters.

\subsection{The Kasner Approximation}
One can see in equations (\ref{confteq}) that a change of the constant $\Lambda$ is equivalent to rescaling the size of the universe in their right-hand sides, or if to abandon the usual range of the Euler angles, it can be compensated by rescaling the spatial coordinates. In the Bianchi I limit, when only the time-derivative terms matter, this freedom is even stronger, and all the equations (\ref{confteq}, \ref{phi}, \ref{constraint}) are then invariant under constant rescalings of $\tau$.

We assume that the right-hand sides of equations (\ref{confteq}) are negligible for some time interval. In this regime,
the equations (\ref{confteq}) reduce to
\begin{equation}
  \alpha_{,\tau\tau} = \beta_{,\tau\tau} = \gamma_{,\tau\tau} = 0,
  \label{eq:kasner_eom}
\end{equation}
whose general solution is
\begin{equation}
  \alpha = p_1\tau + \alpha_0,\quad
  \beta  = p_2\tau + \beta_0,\quad
  \gamma = p_3\tau + \gamma_0,
  \label{eq:kasner_sol_tau}
\end{equation}
with $p_i$, $\alpha_0$, $\beta_0$, $\gamma_0$ being the integration constants.
Combined with the scalar $\phi(\tau) = {\tilde q}\tau + \phi_0$, the solution is characterized by four parameters: $p_1,p_2,p_3,{\tilde q}$.

Note that the only constraint on the parameters comes from the equation (\ref{constraint}). One can correctly say that this is too much freedom for the familiar Bianchi I solutions. This is true, and it comes from the remnant freedom of diffeomorphisms. Every value of $\Lambda\neq 0$ is good enough, and it changes all the $\tau$ derivatives by a factor of $\Lambda$, with no change to the equations. In particular, if we take a solution of
$$a(\tau) = e^{p_1 \tau}, \qquad b(\tau) = e^{p_2 \tau}, \qquad c(\tau) = e^{p_3 \tau} \qquad \mathrm{with} \qquad \Lambda t = \frac{e^{(p_1 + p_2 + p_3)\tau}}{p_1 + p_2 + p_3}$$
according to the transformation (\ref{tau}), then in the physical time it means
$$a(t) = \left(\Lambda \left(\sum p_i\right) t\right)^{\frac{p_1}{\sum p_i}}, \qquad b(t) = \left(\Lambda \left(\sum p_i\right) t\right)^{\frac{p_2}{\sum p_i}}, \qquad c(t) = \left(\Lambda \left(\sum p_i\right) t\right)^{\frac{p_3}{\sum p_i}}$$
so that only the combinations of 
$${\tilde p}_i \equiv \frac{p_i}{p_1+p_2+p_3}$$ 
are the real Kasner exponents.

Therefore, one should better choose a solution with the normalization condition of
\begin{equation}
  p_1 + p_2 + p_3 = 1,
  \label{eq:kasner_sum}
\end{equation}
so that it directly corresponds to the Kasner exponents, $a\sim t^{p_1}$, $b\sim t^{p_2}$, $c\sim t^{p_3}$ in the physical time. It does not restrict our choice of a physical solution in this regime, and if it comes as a limit of an exact solution, one can renormalize the value of $p_1 + p_2 + p_3 $ by a simple rescaling of the time variable $\tau$.

Given the normalization (\ref{eq:kasner_sum}), the equation (\ref{constraint})
$$2(p_1 p_2 + p_1 p_3 + p_2 p_3) = \epsilon\frac{{\tilde q}2}{2}$$
 takes the form of
\begin{equation}
  p_1^2+p_2^2+p_3^2 + \epsilon\frac{{\tilde q}^2}{2}
  = 1.
\end{equation}
Defining $q\equiv {\tilde q}/\sqrt{2}$, this reads
\begin{equation}
  p_1^2 + p_2^2 + p_3^2 +\epsilon q^2 = 1,
  \label{eq:kasner_q}
\end{equation}
and we will be interested in the case of $\epsilon=-1$, or $p_1^2 + p_2^2 + p_3^2 = 1+q^2$.

Even though, in any prechosen time $\tau$, the variable $q$ is constant, we must change the time after a bounce for it to correspond to the physical Bianchi I parameters. Namely, we will need such $\tau$ for which the condition (\ref{eq:kasner_sum}) holds again. Or, for the numerical simulations, one might define
\begin{equation}
\label{accdef}
p_1 = \frac{\alpha_{, \tau}}{\alpha_{, \tau} + \beta_{,\tau} + \gamma_{\tau}}, \qquad p_2 = \frac{\beta_{, \tau}}{\alpha_{, \tau} + \beta_{,\tau} + \gamma_{\tau}}, \qquad p_3 = \frac{\gamma_{, \tau}}{\alpha_{, \tau} + \beta_{,\tau} + \gamma_{\tau}}, \qquad q= \frac{\phi_{, \tau}/\sqrt{2}}{\alpha_{, \tau} + \beta_{,\tau} + \gamma_{\tau}}.
\end{equation}
 Setting $q=0$ recovers the standard vacuum Kasner with $\sum p_i=1$ and $\sum p_i^2=1$.

\subsection{A bounce}
 
Now let us assume that one of the exponents is negative, say $p_1 < 0$. The corresponding scale factor will drive a bounce. Note that, unlike in vacumm, this is just an assumption. For a canonical scalar field, there might be a situation with no negative $p_i$. For a phantom field, two negative exponents are possible.

Keeping only the $\alpha$-terms on the right-hand sides of equations (\ref{confteq}) with $\Lambda=1$, we get
\begin{equation}
\label{eq:bounce_eqs}
  \alpha_{,\tau\tau} = -\tfrac{1}{2}e^{4\alpha},
  \qquad
  \beta_{,\tau\tau}   = \tfrac{1}{2}e^{4\alpha},
  \qquad
  \gamma_{,\tau\tau}  = \tfrac{1}{2}e^{4\alpha}.
\end{equation}
The first equation in (\ref{eq:bounce_eqs}) is a particle motion in an exponential potential, with the energy conservation law
\begin{equation}
  \alpha_{,\tau}^2 + \tfrac{1}{4}e^{4\alpha} = E = \mathrm{const}.
  \label{energy_alpha}
\end{equation}
Having hit the potential wall, the particle moves back and, in the limit of the potential energy negligible again, it comes to 
$$\alpha_{, \tau}=-p_1,$$ the same kinetic energy as it had with $+p_1$.

Other equations (\ref{eq:bounce_eqs}) then show that $\alpha_{,\tau} + \beta_{,\tau}$ and $\alpha_{,\tau} + \gamma_{,\tau}$ are constant. Therefore, after the bounce, when one can neglect the difference between $\alpha_{,\tau}$ and $-p_1$, we have
$$-p_1 + \beta_{,\tau} = p_1 + p_2, \qquad -p_1 + \gamma_{,\tau} = p_1 + p_3.$$
Altogether, it means that
\begin{equation}
\label{newvol}
\alpha_{,\tau} + \beta_{,\tau} + \gamma_{,\tau} = -p_1 + (p_2+2p_1) + (p_3+2p_1) = 1 + 2p_1
\end{equation}
if the normalization condition (\ref{eq:kasner_sum}) was satisfied before the bounce. Therefore, we must either rescale the time $\tau$ by the factor of $1+2p_1$ or use the accurate definitions (\ref{accdef}) for the new Kasner parameters. In any case, we get
\begin{equation}
\label{1bounce}
p_1' = \frac{-p_1}{1+2p_1},
\qquad
  p_2' = \frac{p_2 +2p_1}{1+2p_1},
  \qquad
  p_3' = \frac{p_3 +2p_1}{1+2p_1},
 \qquad
  q'   = \frac{q}{1+2 p_1}.
\end{equation}

In relation to the discussion above, let us note again that the Kasner exponents are unambiguously defined (in the physical time) for the Kasner I solutions only. In order to keep track of them outside the Kasner I epochs, we introduce the definitions (\ref{accdef}). Note also that, in a general Bianchi IX dynamics with such a definition,
$q$ is no longer constant.

\section{Two negative exponents}

An important difference from the standard BKL is that, with a phantom, there might be two negative exponents, $p_1\leqslant p_2<0$. That means that there are two dimensions that grow towards the singularity in the past. One would correctly expect that, if such is the case, the bounce will most probably occur driven by a dimension with the more negative exponent. However, it might also be another negative exponent in case of severe difference in the initial sizes, as well as more difficult dynamics when both dimentions do play a role, especially if the two exponents are close to each other.

Having neglected the third dimension and put $\Lambda=1$, we get the Bianchi IX equations (\ref{confteq})
\begin{equation}
\label{twonegeq}
2\alpha_{,\tau\tau} = e^{4\beta} - e^{4\alpha}, \qquad 
    2\beta_{,\tau\tau} = e^{4\alpha} - e^{4\beta}, \qquad
    2\gamma_{,\tau\tau} = (e^{2\alpha} - e^{2\beta})^2
\end{equation}
as well as the scalar field motion (\ref{phi}) and the constraint (\ref{constraint}) as before,
\begin{equation}
\label{twonegscal}
\phi_{,\tau\tau} = 0, \qquad 2\bigl[\alpha_{,\tau}\beta_{,\tau} + \alpha_{,\tau}\gamma_{,\tau} + \beta_{,\tau}\gamma_{,\tau}\bigr] = \alpha_{,\tau\tau} + \beta_{,\tau\tau} + \gamma_{,\tau\tau} - \frac{1}{2}(\phi_{,\tau})^2.
\end{equation}
From the equations (\ref{twonegeq}, \ref{twonegscal}) we have $\alpha_{,\tau\tau} + \beta_{,\tau\tau} =0$, and can choose the origin of time such that
\begin{equation}
\alpha + \beta = p_{+}\tau, \label{eq:sum}
\end{equation}
where $p_1+p_2=p_{+} <0$ is a constant.

Now, we give a name to the difference of $\alpha$ and $\beta$, and introduce yet another time variable $\eta$: 
$$\chi \equiv \alpha - \beta, \qquad \ln (-2p_+ \eta) =p_{+}\tau $$  
so that 
\begin{equation}
\label{absol}
     2\alpha =\ln(-2p_+ \eta)+\chi,  \qquad 2\beta =\ln(-2p_+ \eta)-\chi.  
\end{equation}
Note that the times $\eta$ and $\tau$ go in opposite directions. By subtracting the equations (\ref{twonegeq}) for $\alpha$ and $\beta$ from each other, one can get the following equation for $\chi$:
\begin{equation}
    \chi_{,\eta\eta} +\frac{\chi_{,\eta}}{\eta} +  \frac12\sinh(2\chi) = 0 \label{1}.
\end{equation}
Substituting the $\gamma_{,\tau\tau}$ expression (\ref{twonegeq}) and solutions (\ref{absol}) for $\alpha$ and $\beta$ into the scalar constraint equation (\ref{twonegscal}), with subsequent transition from $\tau$-derivatives to $\eta$-derivatives, results in
\begin{equation}
    4\,\frac{\gamma_{,\eta}}{\eta} = -\frac{Q}{\eta^2} + \chi_{,\eta}^2 + \frac12 \left(\cosh 2\chi -1\right).
    \label{2}
\end{equation}
where $Q\equiv 1+ \frac{{\tilde q}^2}{p_+^2}$ and $\phi_{,\tau}=\tilde q$ as before.

Solving these equations is not a simple task. This is why we have investigated the problem using both numerical and analytic techniques. But before we study the solution analytically in certain limiting cases, it is instructive to try to understand the behavior of the function $\chi$. After multiplying by \(\chi_\eta\), equation (\ref{1}) can be rewritten as
$$
\frac{d}{d\eta} \left( \frac{\chi_{,\eta}^{2}}{2} + \frac14 \cosh 2\chi \right) = -\frac{1}{\eta} \chi_{,\eta}^{2}.
$$
This equation describes the motion of a particle in a potential \(U =\frac14 \cosh 2\chi\), with decreasing total energy due to friction, with friction coefficient $\sim$ \(1/\eta\). 

\subsection{Case i: $\chi \rightarrow \infty$ and Case ii: $\chi \rightarrow -\infty$, and a general case}

For $|\chi| \gg 1$, it is easier to solve the system (\ref{twonegeq}, \ref{twonegscal}) using the $\tau$-time. However, it is also enough to recall the case of one negative exponent. Indeed, in this section, we neglect the contribution of $c$ to the right-hand side of equations, while a huge value of $|\chi|$ means that one of the remaining two scale factors can also be neglected.

Therefore, in {\bf Case i} of $\chi \gg 1$, we get $a \gg b$ and, given also that $p_1 < p_2$, we are fully back to the case of one negative exponent (\ref{1bounce}). 

In {\bf Case ii} of $\chi \ll -1$, we get $b \gg a$ and, if the inequality is strong enough to initiate a bounce before the factor $a$ takes over due to $p_2 > p_1$, then we get a bounce driven by the dimension $b$ only, with its exponent $p_2$. Since the map (\ref{1bounce}) was deduced by assuming that only one dimension was relevant and without any assumptions on the values of other $p_i$-s, the new result is

\begin{equation}
\label{2bounce}
    p_2' = \frac{-p_2}{1+2p_2}, \qquad
    p_1' = \frac{p_1 + 2p_2}{1+2p_2}, \qquad
    p_3' = \frac{p_3 + 2p_2}{1+2p_2}, \qquad
    q' = \frac{q}{1+2p_2}.
\end{equation}
In other words, $\alpha$ and $\beta$ do switch places.

{\bf In general}, if $p_1 \ll p_2 < 0$, we would expect to see a bounce driven by the dimension $a$ of the more negative exponent $p_1$, unless the initial absolute sizes are extremely different from each other in the direction of preferring the slower growing $b$. Hence, the additional information about values of scale factors not encoded into Mixmaster approximation is needed. If the two negative exponents are close to each other, then any of the two options (\ref{1bounce}, \ref{2bounce}) is quite possible, or even more complicated dynamics with two important dimensions. Below we will consider the simplest case of the latter option, the one when the two scale factors, $a$ and $b$, are close to each other over an extended period of time; that is, the variable $\chi$ stays small.

\subsection{Case iii: $|\chi| \ll 1$}

For $|\chi| \ll 1$, the system of equations (\ref{1}, \ref{2}) reduces to
\begin{equation}
\label{twoneg}
    \chi_{,\eta\eta} + \frac{\chi_{,\eta}}{\eta}  = -\chi, \qquad
    \frac{\gamma_{,\eta}}{\eta} =  - \frac{Q}{4\eta^2} + \frac{\chi_{,\eta}^2}{4} + \frac{\chi^2}{4}.
\end{equation}
The first equation (\ref{twoneg}) can be solved exactly as an arbitrary combination of $J_0$ and $Y_0$ Bessel functions.

A possible solution would be
\begin{equation}
\label{nchi}
\chi(\eta) = \frac{\pi (p_1 - p_2)}{2 (p_1 + p_2)}\cdot Y_0 (\eta) + C \cdot J_0 (\eta).
\end{equation}
The (small) coefficient in the first term is chosen so as to reproduce the $\eta\longrightarrow 0$ asymptotic of  
$$ \chi_{,\tau}=p_{+}\, \eta \chi_{,\eta}=p_1-p_2,$$ 
with $\chi_{,\tau}<0$ and $\chi_{,\eta}>0$. Here we have used the $Y_0(\eta) \simeq \dfrac{2}{\pi}\ln\eta + \mathcal{O}(1)$ asymptotic. The second coefficient $C$ can be chosen freely since  $J_0(\eta) \longrightarrow 1$ when $\eta\longrightarrow 0$. In the $\eta\longrightarrow\infty$ limit we then have
\begin{equation}
\label{nchia}
\chi(\eta) = \frac{p_1-p_2}{p_1+p_2}\sqrt{\frac{\pi}{2}}\cdot \frac{\sin{\left(\eta - \frac{\pi}{4}\right)}}{\sqrt{\eta}} + \sqrt{\frac{2}{\pi}}C\cdot \frac{\cos{\left(\eta - \frac{\pi}{4}\right)}}{\sqrt{\eta}} +{\mathcal O}(\eta^{-3/2}).
\end{equation}

An interesting behavior is then seen in the equation (\ref{twoneg}) for $\gamma$. The $\eta\longrightarrow 0$ limit must, of course, correspond to $\gamma_{,\tau}=p_{+}\, \eta \gamma_{,\eta}=p_3$ with $\gamma_{,\tau}>0$ and $\gamma_{,\eta}<0$. In other words, $\gamma$ decreases in time $\eta$. One can also check it explicitly. Indeed, we take the $\eta\longrightarrow 0$ limit of the the function (\ref{nchi}) and substitute it into the formula (\ref{twoneg}) for $\gamma_{,\eta}$:
$$\frac{\gamma_{,\eta}}{\eta} = \frac{-Q+\left(\frac{p_1 - p_2}{p_1+p_2}\right)^2}{4\eta^2} \cdot \left(1 + {\mathcal O}(\eta)\right).$$
Recalling that $Q=1+\frac{2q^2}{(p_1+p_2)^2}$, we calculate the coefficient to be
$$\frac{(p_1-p_2)^2-(p_1+p_2)^2-2q^2}{4(p_1+p_2)^2}=\frac{-2(p_1+p_2)^2-2p_3^2+2}{4(p_1+p_2)^2}=\frac{p_3}{p_1+p_2}$$
where we first have used $q^2=p_1^2+p^2_2+p_3^2-1$ and then, in the enumerator, $(p_1+p_2)^2=(p_1+p_2)(1-p_3)$ and $p_3^2=p_3(1-p_1-p_2)$ as well as $p_1+p_2+p_3=1$.

Therefore, we have checked the correct behavior at the small $\eta$ limit. Then, at finite times, or more precisely, with $\eta$ of order one, we recall that $\chi$ is small, or according to (\ref{nchi}), $|p_1 - p_2|$ and $|C|$ are small. It means that $\frac{\gamma_{,\eta}}{\eta} \approx  - \frac{Q}{4\eta^2}$  if $Q$ is big enough, and therefore we have $\gamma \sim -\frac{Q}{4} \ln\eta$, and $\gamma$ continues to decrease in $\eta$.

However, finally, as the solution (\ref{nchi}) for $\chi$ tends to the asymptotic (\ref{nchia}) of $\eta\longrightarrow\infty$, we see that (for $C=0$)
$$\frac{\gamma_{,\eta}}{\eta} = \frac{\pi(p_1-p_2)^2}{8(p_1+p_2)^2}\cdot \frac{1}{\eta} +{\mathcal O}\left(\frac{1}{\eta^2}\right)$$
which means that $\gamma$ has started growing in time $\eta$, namely $\gamma\propto \eta$. 

As we see, the approximations of this section will inevitably break down at some finite value of time $\eta$ due to the growth of $\gamma$. One can easily see that it also indicates a change in the volume behavior. Indeed, since $\alpha + \beta$ continues to grow (\ref{eq:sum}), with $\chi$ describing oscillations of $\alpha - \beta$, the volume starts growing even a bit before the change in sign of $\gamma_{,\eta}$ (see Fig.1).

\begin{figure}
    \centering
    \includegraphics[width=.49\linewidth]{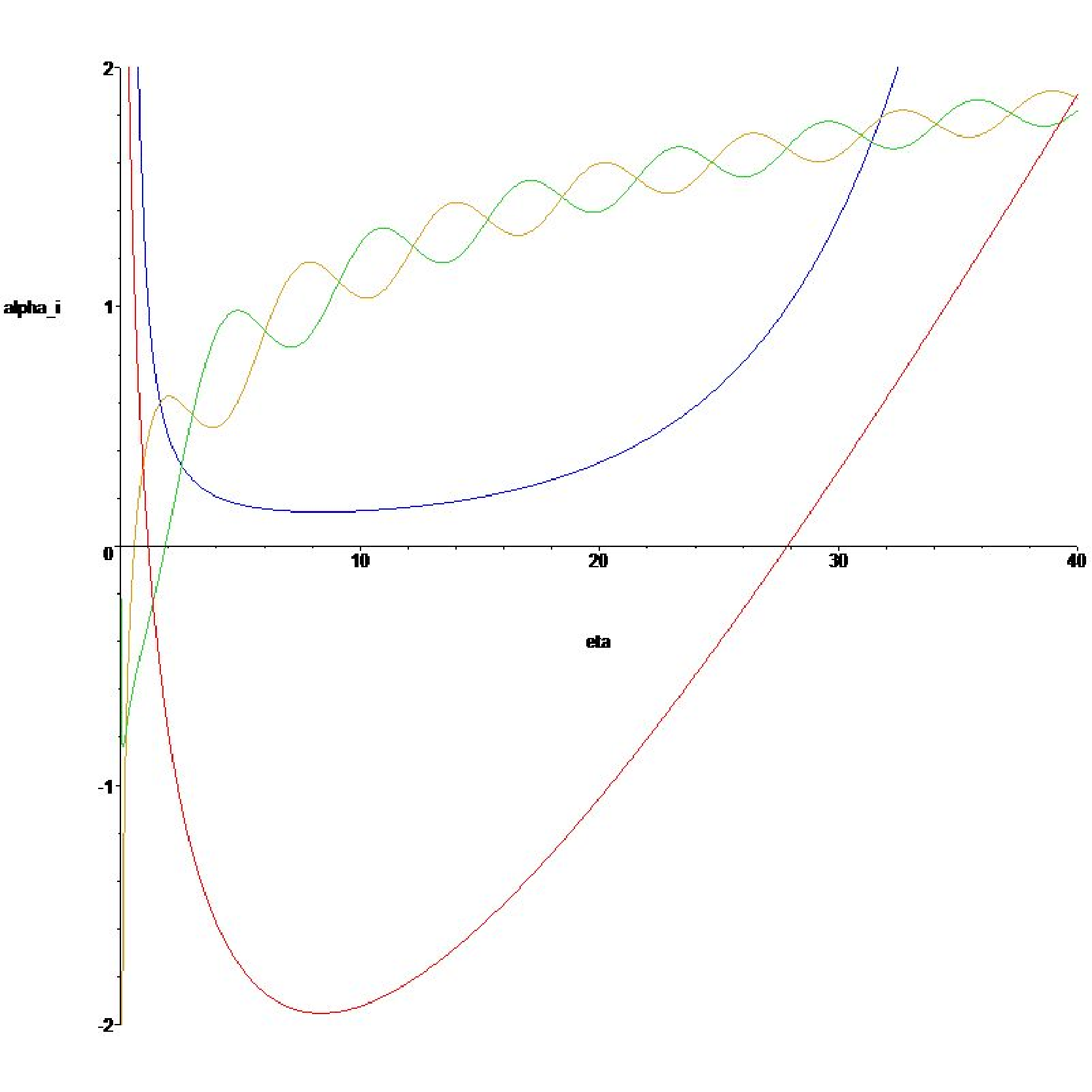}
    \caption{Exponents; $\alpha$, $\beta$, $\gamma$ (yellow, green and red) drawn with the volume (blue) in the case of $\chi\ll 1$. The volume evolution show that the volume rate of change flip sign after certain time, around the time when $\gamma$ flips sign also.}
    \label{fig:frozen_phases}
\end{figure}

\subsection{Summary}

Summarizing, the case of two negative exponents has two important modifications to the standard BKL oscillations. When two corresponding scale factors are very different, the usual formulas for Kasner
exponents hold with the following exception: information on the values of $p_i$ only is now insufficient to predict $p_i'$. Indeed, it is not necessarily the smallest $p_i$ that corresponds to the direction of the potential wall, but the biggest scale factor does. In other words, in order to choose between (23) and (30), we need information on the actual values of the scale factors corresponding to these two negative $p_i$-s. In the standard case, there is only one negative $p$, so this problem does not exist.

In order to avoid switching between two sets of formulas, we can agree to use (23) only, but to $\it define$ $p_1$ as the index corresponding to the scale factor actually inducing the bounce in question,
regardless of its numerical value. In the next Section, we will see that there is yet another aspect of choosing $p_1$ when the volume has changed its behavior.

If the two scale factors are very close to each other, it is more natural to consider not a sequence of Kasner eras, but a continuous evolution (32). This regime is also known for the standard BKL case \cite{BKL2} as small oscillations near the Taub solution (corresponding to $p_1=p_2=0$, $p_3=1$). In contrast to that, now $p_1$ and $p_2$ need not be small. Due to this, the volume derivative
changes its sign near the point of minimum of the 3-rd scale factor.

If two scale factors are exactly equal to each other, the potential wall is absent, and if the 3-rd scale factor is decreasing, there are no bounces at all. If the 3-rd scale factor is increasing, it induces a single bounce with no bounces after that. The solution in both cases gradually approaches 
a locally rotationally symmetric (LRS) Bianchi I solution, determined by the initial value of $\dot \phi$ (see Fig.2).
In the standard BKL picture, the only possible future attractor LRS is the Taub solution.

So far, we have demonstrated the non-monotonic volume behavior for the case of two almost equal scale factors only. Actually, looking at the formula (\ref{newvol}), one can observe this effect even for a single negative $p$ when $p_1 < - \frac12$. In the following section, we will discuss the proper reasons why volume oscillation is a typical regime in the phantom dynamics.

\begin{figure}
\includegraphics[width=.49\linewidth]{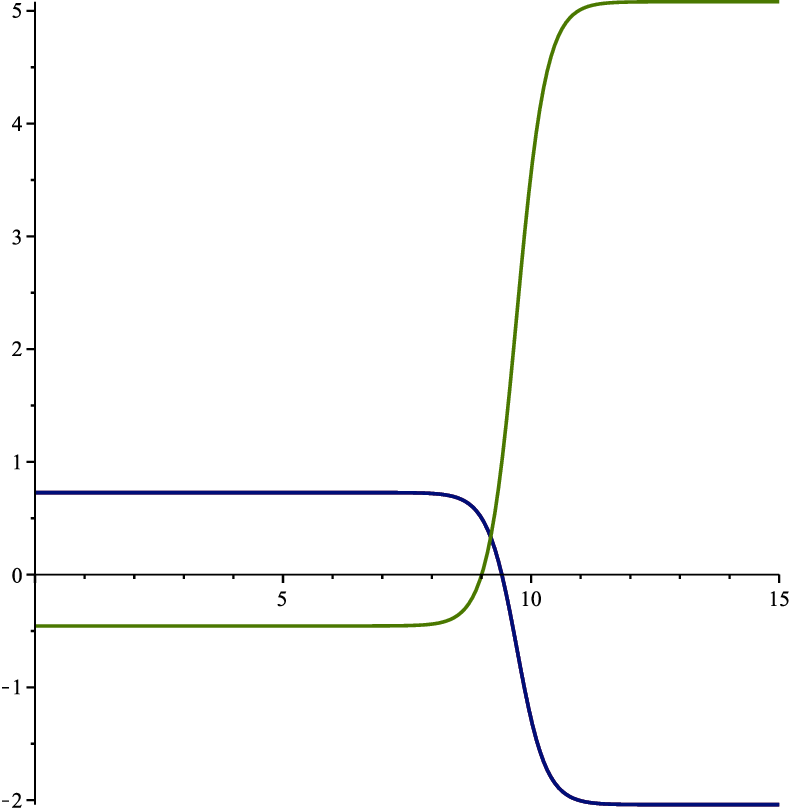}\hfill
\includegraphics[width=.49\linewidth]{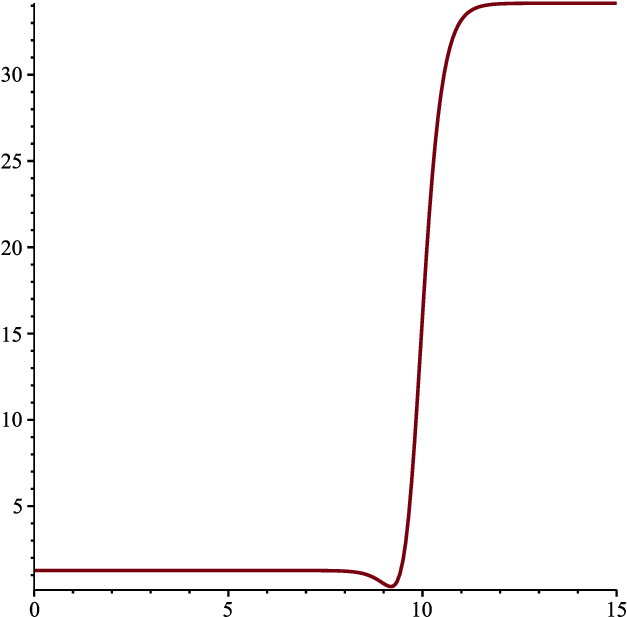}
\caption{Kasner indices (left) and $q$ (right) for an example of  LRS solution.}
\end{figure}

\section {Volume oscillations}
A crucial  feature of the BKL dynamics with a phantom is the fact that the least possible value of Kasner index can become large and negative. It leads to important consequences. To understand them, let us recall the structure of the Mixmaster map (\ref{1bounce}). As after the bounce 
$$\alpha_{,\tau}=-p_1, \qquad \beta_{,\tau}=p_2+2p_1, \qquad \gamma_{,\tau}=p_3+2p_1,$$ 
we see that (\ref{newvol})   
\begin{equation}
\label{volsign}
\frac{dV}{d\tau}=\frac{d(abc)}{d\tau}=V\cdot (1+2p_1).
\end{equation}
Getting the linear dependence $dV/dt = 1+2p_1$ with the standard  equation $p'_1+p'_2+p'_3=1$ after the bounce is achieved by returning to the cosmic time
$dt=abcd\tau$ \cite{BKL2}.

Note, however, that  if $p_1<-1/2$, the volume changes its behavior (\ref{volsign}), and a contraction (when integrating back in time) turns
to expansion. In a vacuum case this situation is impossible since $p_1>-1/3$ while for the phantom case this possibility appears when $q^2>3/8$. Since each bounce with $-1/2 < p_1 < 0$
leads to an increase of $q^2$, this condition can be reached after some finite sequence of bounces. When $q^2$ becomes large enough, it necessarily leads to $p_1 < -1/2$, even though $p_2$ might be small negative and can induce another bounce. We would expect that finally $p_1$ becomes so negative that it must win over anything that is small negative. After that, the overall volume expansion rate changes sign.

At this point point, we can continue using the same definition (\ref{accdef})
that leads to $p_1+p_2+p_3=1$ for both contraction and expansion, or to reverse the signs of $p_i$ when contraction starts. In the next subsection, we will show why the former definition is natural from the viewpoint of correspondence to Bianchi I solutions.

The disadvantage of this choice (\ref{accdef}) is that knowing $p_i$ is not enough to determine the sign of the volume expansion rate, so this information should be added to $p_i$  in order to use the Mixmaster map. On the other hand, this information is reflected in the sign of $q$, and the map itself does not change. For both contraction and expansion we should denote by $p_1$ the index corresponding to the particular scale factor inducing the bounce in question.
This information requires a knowledge of scale factors and, anyway, is not encoded in $p_i$ themselves. 

At the standard BKL stage, the bounce occurs due to curvature terms of a scale factor with a negative index (and, in the phantom case it may be up to 2 such indices, so an additional information is needed), while during the reversed behavior of volume a bounce is induced by a positive index. In other words, one needs to use a positive index as $p_1$ in the map (\ref{1bounce}). This is an important point. Once $p_1$ becomes less than $-\frac12$, the resulting $p_1^{\prime}$ is also negative, and if one uses it as a new $p_1$ (instead of a positive one), then we get back to the previous $p_1$, while the actual dynamics is more complicated.

An {\bf alternative} way, with the change of signs of indices, requires adjusting the Mixmaster map. In this case, all the way until the volume reverts it rate, we need to use
$$p'_1=\frac{-p_1}{|1+2p_1|} \qquad p'_2=\frac{p_2+2p_1}{|1+2p_1|} \qquad p'_3=\frac{p_3+2p_1}{|1+2p_1|}$$
where the modulus sign is important for the last bounce with $p_1<-1/2$ only. After that, one gets a new normalization of $p_1 + p_2 + p_3 = -1$ and changes the map to
$$ p'_1=\frac{-p_1}{|-1+2p_1|} \qquad p'_2=\frac{p_2+2p_1}{|-1+2p_1|} \qquad p'_3=\frac{p_3+2p_1}{|-1+2p_1|}.$$
In both situations, the bounce in this approach is induced by a negative index. We do {\bf not} follow this option here in order to keep the standard Mixmaster map (\ref{1bounce}). 

\begin{figure}[!h]
\includegraphics[width=.32\linewidth]{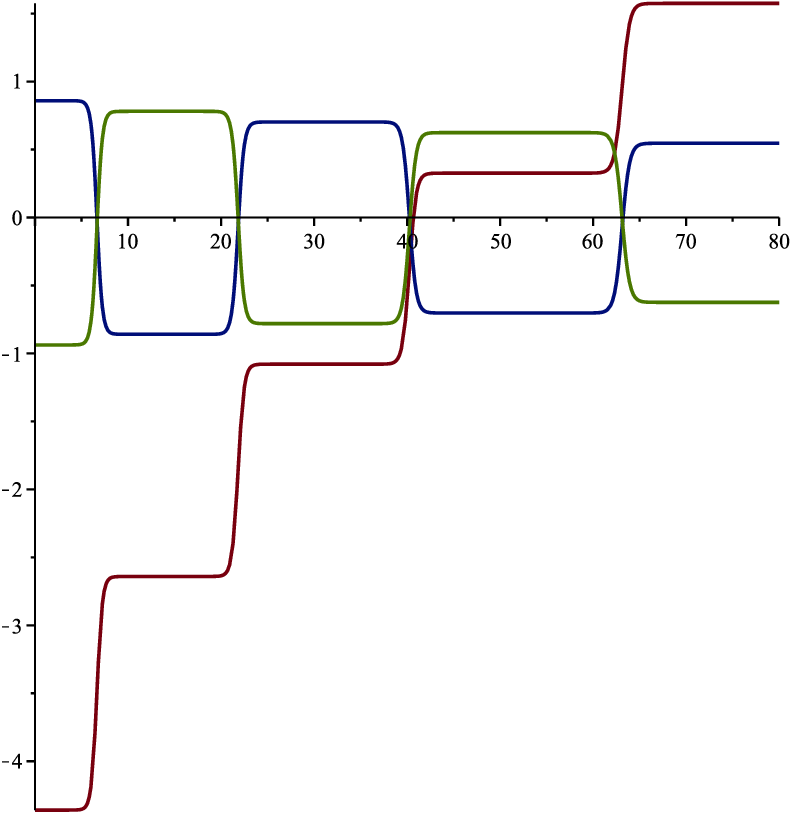}\hfill
\includegraphics[width=.32\linewidth]{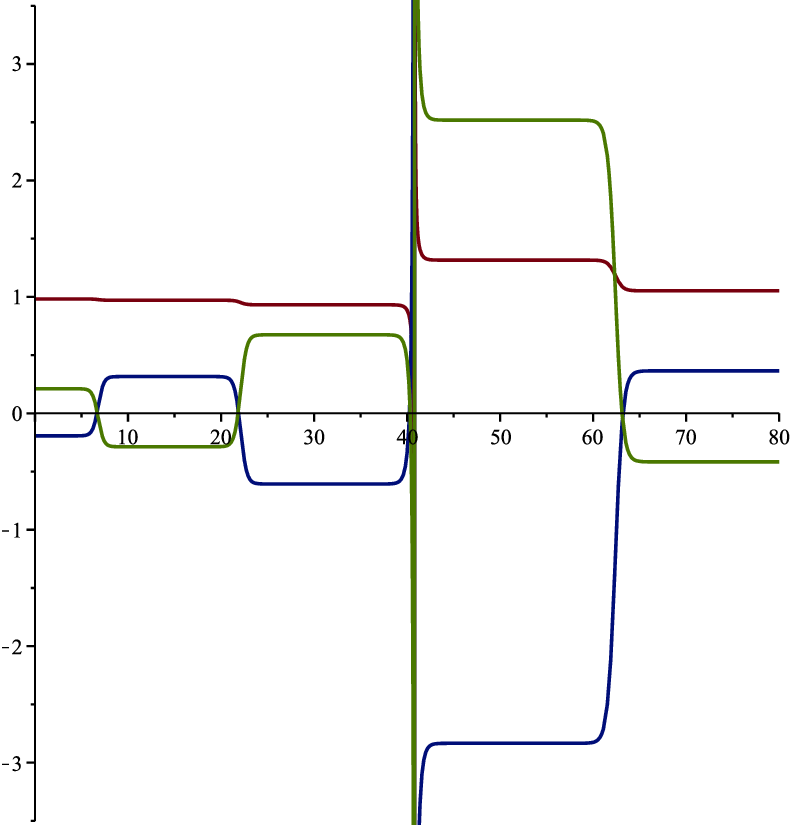} \hfill
\includegraphics[width=.32\linewidth]{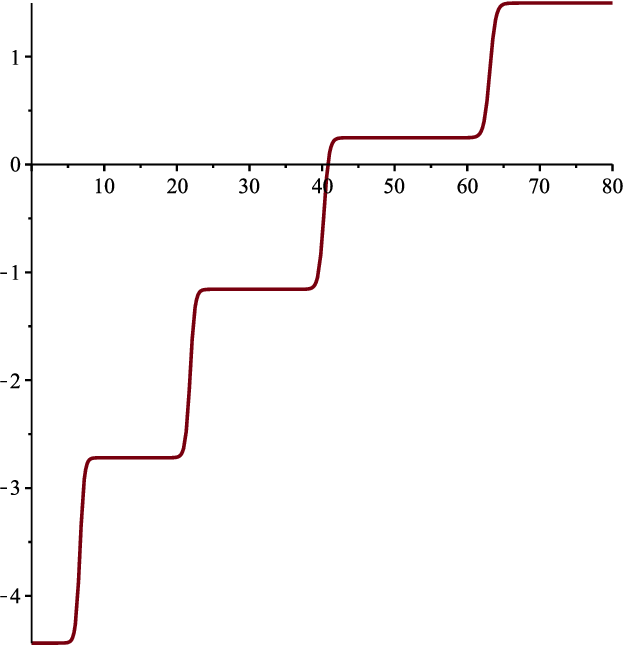}
\caption{ Time evolution of $\alpha,_{\tau}$ (brown), $\beta,_{\tau}$ (blue) and $\gamma,_{\tau}$ (green) -- left panel, corresponding evolution of $p_1$ (brown), $p_2$ (blue), $p_3$ (green) -- central panel,
volume derivative $dV/dt$ -- right panel, for the example 1.}
\end{figure}

In our plots below, we adopt the conservative possibility (\ref{accdef}). Numerical studies confirm the suggested volume behavior. 
A numerical example is shown in Fig.3. We see that after the second bounce $p_1$ drops below $-1/2$ and, as expected, volume derivative changes its sign at the 3-d bounce.
In Fig.4 we present a situation where initially $p_1$ is very close to $-1/2$. We see that evolution of all variables except for $p_i$ is smooth, and the Kasner
epochs are clearly seen as time intervals when scale factor derivatives  are almost constant. As for $p_i$, they are discontinuous at the first bounce and get very large values ($|p_i|$ 
are of the order of hundred for this particular example) right after that, so we do not plot them here, showing the behavior of $\alpha$, $\beta$ and $\gamma$ instead.
Such behavior does not spoil smoothness of other variables.
As expected, after the second bounce the volume starts to increase while integrating into the past. At the 3-rd bounce we can detect a situation mentioned  above - two indices are positive, and we need additional information about the nature of the bounce. In our example, it appears that  the actual
bounce is experienced by a scale factor corresponding to smaller value of the index. So that, despite
the index corresponding to $\beta$ being the second one by ordering, it should be denoted as $p_1$ for the correct calculation of the indices after the 3-rd bounce.

\begin{figure}
\includegraphics[width=.32\linewidth]{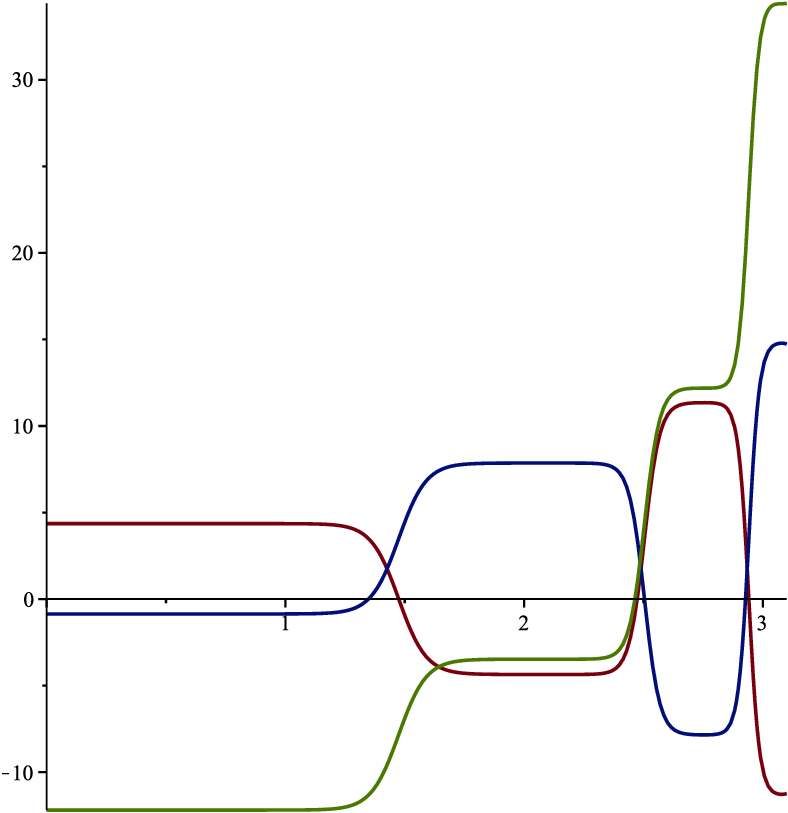}\hfill
\includegraphics[width=.32\linewidth]{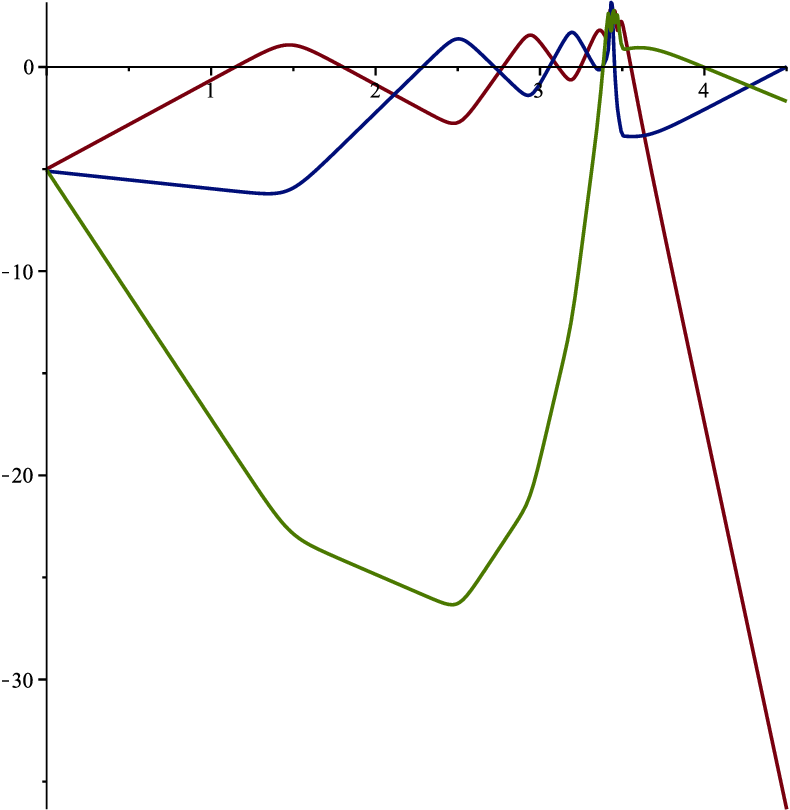} \hfill
\includegraphics[width=.32\linewidth]{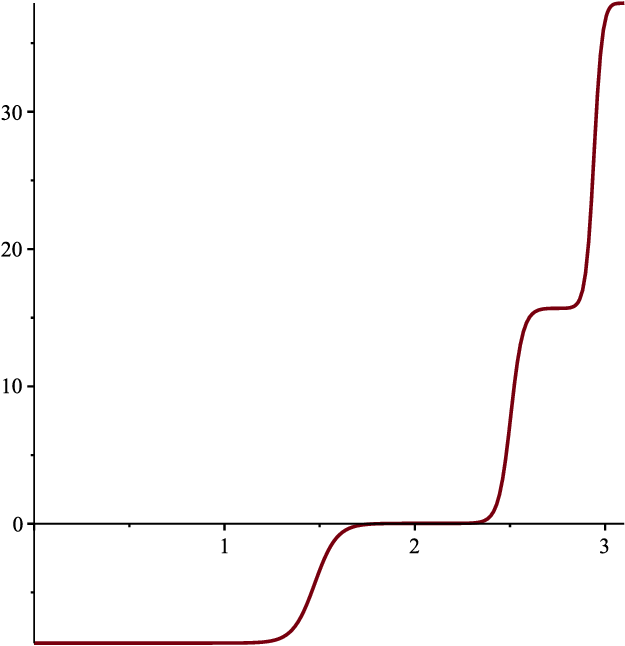}
\caption{ Time evolution of $\alpha,_{\tau}$ (brown), $\beta,_{\tau}$ (blue) and $\gamma,_{\tau}$ (green) -- left panel, corresponding evolution of $\alpha$ (brown), $\beta$ (blue) and $\gamma$ (green) -- central panel,
volume derivative $dV/dt$ -- right panel, for the example 2.}
\end{figure}

In the paper \cite{Volkov} it was claimed that all numerical integrations provided by the author indicate the absence of a singularity, since the volume starts to increase at some point. Our 
numerical results show the same. Though a strict mathematical prove of the absence of a cosmological singularity in Bianchi IX Universe with a phantom field is still to be constructed, the connection 
between a volume bounce and the condition $p_1<-1/2$ makes this conjecture to be at least reasonable to expect as we have mentioned above. 

Indeed,
we have our Kasner exponents with $\sum p = 1$ and $\sum p^2 > 1$. The difference from 1 in the last inequality is due to the negative energy density. After a scale factor bounce this energy  becomes even more important  since, as long as $-1/2 < p_1 < 0$, the denominator $1 + 2p_1$ in the transformation law for $q$ is small and positive. Therefore, the scalar field continues to behave the same way as before, but with a larger value of $q^2$. In other words, $\sum p^2$ becomes even larger. It is easy to see that for the combination of one negative and two positive indices $p_1$ ultimately drops below $-1/2$
for any $q^2>3/2$ (which leads to a volume bounce),   and the least $p$ is less than $-1/2$ for the two negative indices case if $q^2>7/2$. However, this latter case needs more investigation since a bounce can be induced by the second 
$p$ (if we use the standard ordering) which can be close to zero even for large $q$.

 It is easy to see that the transition from expansion to contraction (which can be seen, for example,  in Fig.4 for the time near $3.5$) can not be explained the same way.
Indeed, for our definition of $p_i$ the denominator of Mixmaster map at expansion can not be negative since the bounce is induced by a positive index.
However, it is known that even a vacuum Bianchi IX Universe must have a point of maximal volume, and after that a collapse starts \cite{Wald1, Wald2}. The same situation appears in the phantom case also,
and near this point the Mixmaster approximation breaks down. 

This is to be expected. Indeed, the spatial topology of the Bianchi IX universe is that of a sphere. In terms of an isotropic background, both the anisotropies and the massless phantom behave as $\frac{1}{a^6}$, where $a$ is the mean scale factor, while the (positive) isotropic curvature goes as $\frac{1}{a^2}$, and the latter must finally come to domination and cause recollapsing. At the same time, a key point of the Mixmaster description is in the approximate Bianchi I epochs. Those behave in a purely $\frac{1}{a^6}$ manner, and that must break down at the really large volume.

So, we see that volume extrema have different origin. While volume maximum is typical for Bianchi IX evolution for  the vacuum case as well, the volume minima appear only due to the presence of phantom field.
As the points of volume local maximum are inevitable, and the Mixmaster map fails in the vicinity
of such a point, this means that it makes no sense to iterate the Mixmaster map infinitely.
Moreover, the isotropic curvature term $1/a^2$  becomes negligible with respect to the anisotropic one near a singularity, but as the volume in the phantom case oscillates between
some finite values, this term always has finite (though possibly very small) contribution.
We can conclude that for a phantom case the Mixmaster map  is a transient regime realized with some finite accuracy
and {\it it cannot be considered as an attractor for a general Bianchi IX evolution}

\subsection {The Kasner indices of Bianchi I and the case of $p_1 = -\frac12$}
A question may arise: what happens when $p_1=-1/2$ exactly? In principle, this question may be avoided by remarking that in the full Bianchi IX picture the meaning of Kasner exponents is a bit fuzzy, so there is no sense to equate them to any exact number. However, such a value indicates the existence of a new regime in Bianchi I model, specific only to phantom. To describe it in a regular manner, we need to go back to initial variables.

Let us consider a Bianchi I universe given by
$$ds^2 = dt^2 - a^2_i (t) dx_i^2$$
with the Hubble parameters $H_i \equiv \frac{{\dot a}_i}{a_i}$. We will also denote the rate of the volume change as
\begin{equation}
\label{volrat}
{\mathfrak V}\equiv \frac{\dot V}{V} =\sum H_i.
\end{equation}

All the subsequent discussion can be held in any dimension, but we take the physical case of $(1+3)$D. Having put $8\pi G =1$, we get the Friedmann equations
\begin{equation}
\label{Fr0}
H_1 H_2 + H_1 H_3 + H_2 H_3 = \rho,
\end{equation}
\begin{equation}
\label{Fr1}
{\dot H}_2 + {\dot H}_3 + H^2_2 + H_3^2 + H_2 H_3 = - P_1,
\end{equation}
\begin{equation}
\label{Fr2}
{\dot H}_1 + {\dot H}_3 + H^2_1 + H_3^2 + H_1 H_3 = - P_2,
\end{equation}
\begin{equation}
\label{Fr3}
{\dot H}_1 + {\dot H}_2 + H^2_1 + H_2^2 + H_1 H_2 = - P_3,
\end{equation}
with the energy density $\rho$ and pressures $P_i$.

Now, assume that the matter content is a stiff ($w=1$) isotropic fluid, i.e.
$$\rho=P_1=P_2=P_3.$$
By adding any two dynamical equations (\ref{Fr1}, \ref{Fr2}, \ref{Fr3}) to each other, subtracting the third one from them, and adding the constraint equation (\ref{Fr0}) to the result, we get the very simple form of dynamical equations
\begin{equation}
\label{Idyn}
{\dot H}_1 + H_1 {\mathfrak V} =0, \qquad {\dot H}_2 + H_2 {\mathfrak V} =0, \qquad {\dot H}_3 + H_3 {\mathfrak V} =0.
\end{equation}
It is absolutely the same as in vacuum, with the only difference in the constraint equation (\ref{Fr0}).

Note that the equations (\ref{Idyn}) do not depend on any assumption beyond the geometry of Bianchi I and the stiff matter fluid. Namely, they do not depend on whether, on average, the universe is expanding or contracting. Therefore, by adding them all together, we find that
\begin{equation}
\label{volrateq}
{\dot{\mathfrak V}} + {\mathfrak V}^2 =0, \qquad \mathrm{hence} \qquad {\mathfrak V}=\frac{1}{t - t_0}
\end{equation}
always, and we can choose $t_0 =0$ with no lack of generality.

Substituting $ {\mathfrak V}=\frac{1}{t}$ into the equations for $H_i$, one obtains
\begin{equation}
\label{Kasnerind}
H_i = \frac{p_i}{t} \qquad \mathrm{with} \qquad p_1+p_2+p_3=1.
\end{equation}
This is how we get the definition of Kasner exponents, for both expanding and contracting universes:
\begin{equation}
\label{Kasnerindef}
p_i = \frac{H_i}{\mathfrak V} \qquad \mathrm{or} \qquad p_i = - \frac{H_i^2}{\dot H_i},
\end{equation}
and both of them do not care about time reversal. We have taken the first of this formulas (\ref{Kasnerindef}) as our definition (\ref{accdef}) for the indices in the Bainchi IX case.

In complex numbers, the final solution is $a_i(t) \propto t^{p_i}$ for all $t\neq 0$. In real numbers, we need to consider two series of solutions, of increasing and decreasing volumes respectively,
$$a_i(t) \propto t^{p_i} \quad {\mathrm for \ } t>0 \quad \mathrm{and} \quad a_i(t) \propto (-t)^{p_i} \quad \mathrm{for \ } t<0.$$
In particular, a shrinking (${\dot{\mathfrak V}}<0$) universe is the latter option, and it goes as
$$ds^2 = dt^2 - c_i (-t)^{p_i} dx_i^2,$$
starting from infinite volume at negative infinity in time and finishing with a singularity at $t=0$, with $p_1+p_2+p_3=1$, though with $p_i>0$ meaning contraction while $p_i<0$ meaning expansion. One can take it as a justification for the conservative definition (\ref{accdef}) of the Kasner indices.

Note also that the equation for the volume rate of change (\ref{volrateq}) does also have a solution of
\begin{equation}
\label{constV}
{\mathfrak V}=0.
\end{equation}
Normally, it would only correspond to a Minkowski solution in vacuum. However, the negative energy of a phantom allows for an exponential (in physical time) solution. 

Since this is a new general feature, let us illustrate it in arbitrary dimension. Given ${\mathfrak V}=0$, we get (\ref{Idyn})
\begin{equation}
\label{nssol}
{\dot H}_i = 0,
\end{equation}
that is an exponential solution, indeed. The temporal component of Einstein equations (\ref{Fr0}) then takes the form of
\begin{equation}
\label{constener}
\sum_{i,j:\ i<j} H_i H_j =\rho.
\end{equation}
The condition of constant volume (\ref{constV}) in ($1+d$) dimensions means that
$$H_d = - \sum_{i:\ i<d} H_i$$
and finally the constant phantom energy density (\ref{constener}) takes the value of
\begin{equation}
\rho = - \left(\sum_{i:\ i<d}H_i^2 + \sum_{i,j:\  i<j<d} H_i H_j \right) = - \frac12 \left(\left(\sum_{i:\ i<d} H_i\right)^2 + \sum_{i:\ i<d} H_i^2\right) \leqslant 0,
\end{equation}
with the equality for the vacuum Minkowski space only.

Obviously, these non-singular solutions (\ref{nssol}) feature infinite Kasner exponents, if to take the formal definition (\ref{Kasnerindef}). They explain the singularity of the Kasner map (\ref{1bounce}). It corresponds to a measure zero of initial data, of course. However, a new feature of the Mixmaster dynamics is that the Kasner indices can easily become arbitrarily large.

\section*{Conclusions}
In the present paper we have considered the picture of BKL oscillations and explained the numerical results 
of the recent paper \cite{Volkov} showing volume oscillations.

The presence of a phantom field induces several  important modifications of the standard BKL scenario.
As it was mentioned in \cite{Volkov}, a new feature is the fact that two of the three Kasner indices
can be negative. We studied the impact of this possibility in Sec.2. The first consequence is that now
the information encoded in the Mixmaster map is not enough to trace evolution of the full system since
a bounce can be induced by the curvature terms from both scale factors corresponding to negative indices. In order to determine which scale factor is responsible for a bounce, it is necessary to use information about actual values of scale factors, and the information about power indices only is not enough.

Note, however, that a similar feature exists in the standard BKL case when one evolves the system to the future, i.e. away from the singularity, instead of the past (this apparent time asymmetry is mentioned, for example, in \cite{Uggla}). The phantom case indicates clearly that the number of possible outcomes of a considered particular bounce (which is equal to 2 for future and 1 for the past evolution in the standard case) is not generally connected  to the time arrow.

The second effect of two negative indices is that when corresponding scale factors are close to each other, the potential walls are no longer exponential, and instead of the sequence of Kasner epochs we have a smooth evolution.  This regime is not completely new since it has an analog in the standard case - the regime of small oscillations near the Taub solution. However, for the phantom case, it is accompanied
with the volume expansion rate changing its sign.

We show also that two negative indices are not necessary to get the volume extrema.
 We  argue that they appear when the least index drops below $-1/2$, which is impossible without a phantom. After that, the denominator of the standard Mixmaster map (22) changes its sign.
Our numerical studies confirms that after that the volume expansion rate also changes its sign.
Since volume maxima is known to exist even in the vacuum case, these two facts lead to volume oscillations detected in \cite{Volkov}.
So that, keeping the physical explanation of avoiding a cosmological singularity by adding a negative phantom energy presented in \cite{Volkov}, we add a formal argumentation which uses only mathematical properties of the Bianchi IX equations of motion. 

\subsection*{Acknowledgements}
The work of DC and AT was conducted under the state assignment of Lomonosov Moscow State University. AT
is grateful to the British University in Egypt (Cairo), where this work has been initiated, for hospitality. The work of AA is partially supported by the Science, Technology \& Innovation Funding Authority (STDF) under grant number 50806.

\end{document}